\documentclass[10pt,conference,twocolumn,]{IEEEtran}
\IEEEoverridecommandlockouts

\usepackage{amsmath,amssymb,amsfonts}
\usepackage{cite}

\usepackage{graphicx}
\usepackage{textcomp}
\usepackage{xcolor}
\usepackage{times}
\usepackage{subfigure}         
\usepackage{amssymb,amsmath}
\usepackage{acronym}  
\usepackage{balance}

\usepackage{bm}         
\usepackage{algorithm} 
\usepackage{algorithmic}

\usepackage{lipsum}
\usepackage{stfloats}

\allowdisplaybreaks[4]

\newtheorem{proposition}{\bf Proposition}

\acrodef{ofdm}[OFDM]{orthogonal frequency division multiplexing}%
\acrodef{miso-ofdm}[MISO-OFDM]{multi-input single-output orthogonal frequency division multiplexing}%

\acrodef{ris}[RIS]{reconfigurable intelligent surface}%
\acrodef{qos}[QoS]{quality of service}%
\acrodef{idft}[IDFT]{inverse discrete Fourier transform}%
\acrodef{dft}[DFT]{discrete Fourier transform}%
\acrodef{cp}[CP]{cyclic prefix}%
\acrodef{csi}[CSI]{channel state information}%
\acrodef{awgn}[AWGN]{additive white Gaussion noise}%

\acrodef{qcqp}[QCQP]{quadratically constrained quadratic program}%
\acrodef{qp}[QP]{quadratic program}%
\acrodef{bs}[BS]{base station}%
\acrodef{ap}[BS]{base station}%
\acrodef{aps}[APs]{access points}%
\acrodef{qos}[QoS]{quality of service}%
\acrodef{ue}[UE]{user equipment}%
\acrodef{snr}[SNR]{signal-to-noise ratio}%
\acrodef{mmwave}[mmWave]{millimeter-wave}%
\acrodef{snr}[SNR]{signal-to-noise ratio}%

\acrodef{sinr}[SINR]{signal-to-interference-plus-noise ratio}%

\acrodef{ser}[SER]{symbol error rate}%
\acrodef{rc}[RC]{reflection coefficient}%
\acrodef{uavs}[UAVs]{unmanned aerial vehicles}%
\acrodef{mimo}[MIMO]{multiple-input multiple-output}%
\acrodef{noma}[NOMA]{non-orthogonal multiple acces}%

\acrodef{ace}[ACE]{adaptive cross-entropy}%
\acrodef{wsr}[WSR]{weighted sum-rate}%
\acrodef{udn}[UDN]{ultra-dense network}%
\acrodef{Udn}[UDN]{Ultra-dense network}%

\def\BibTeX{{\rm B\kern-.05em{\sc i\kern-.025em b}\kern-.08em
    T\kern-.1667em\lower.7ex\hbox{E}\kern-.125emX}}

\begin{document}
\title{Capacity Improvement in Wideband\\ Reconfigurable Intelligent Surface-Aided \\Cell-Free Network}
\author{
	\vspace{0.2cm}
	\IEEEauthorblockN{
		Zijian~Zhang
		and
		Linglong~Dai}
	\IEEEauthorblockA{\IEEEauthorrefmark{0}
		Beijing National Research Center for Information Science and Technology (BNRist)\\
		Department of Electronic Engineering,
		Tsinghua University,
		Beijing 100084, China\\
		Email: zhangzij15@mails.tsinghua.edu.cn, daill@tsinghua.edu.cn
	}
}
\maketitle
\begin{abstract}
Thanks to the strong ability against the inter-cell interference, cell-free network has been considered as a promising technique to improve the network capacity of future wireless systems. However, for further capacity enhancement, it requires to deploy more base stations (BSs) with high cost and power consumption. To address the issue, inspired by the recently proposed technique called reconfigurable intelligent surface (RIS), we propose the concept of RIS-aided cell-free network to improve the network capacity with low cost and power consumption. Then, for the proposed RIS-aided cell-free network in the typical wideband scenario, we formulate the joint precoding design problem at the BSs and RISs to maximize the network capacity. Due to the non-convexity and high complexity of the formulated problem, we develop an alternating optimization algorithm to solve this challenging problem. Note that most of the considered scenarios in existing works are special cases of the general scenario in this paper, and the proposed joint precoding framework can also serve as a general solution to maximize the capacity in most of existing RIS-aided scenarios. Finally, simulation results verify that, compared with the conventional cell-free network, the network capacity of the proposed scheme can be improved significantly.
\end{abstract}
\section{Introduction}
Network technique is the most essential technique to increase the capacity of wireless communication systems. Recently, a novel \textit{user-centric} network paradigm called cell-free network has been proposed \cite{Nayebi'15} to address the issue of inter-cell interference, which is common in conventional cellular networks. Due to the efficient cooperation among all distributed \ac{bs}s in the cell-free network, the inter-cell interference can be effectively alleviated, and thus the network capacity can be accordingly increased. This promising technique has been considered as a potential candidate for future communication system \cite{Ngo'17}, and has attracted the increasing research interest in recent years.
\par
However, to improve the network capacity further, the deployment of more distributed synchronized \ac{bs}s requires high cost and power consumption in the cell-free network. Fortunately, the emerging new technique called \ac{ris} is able to provide an energy-efficient alternative to enhance the network capacity. Verified by a communication prototype \cite{LinglongDai}, with a large number of low-cost passive elements, \ac{ris} is able to reflect the elertromagentic incident signals to any directions with extra high array gains by adjusting the phase shifts of its elements. Since the wireless environment can be effectively manipulated with low cost and energy consumption \cite{Huang'18'2}, \ac{ris} can be used to improve the channel capacity, reduce the transmit power, enhance the transmission reliability.
\par
In this paper, we first propose the concept of RIS-aided cell-free network to further improve the network capacity of the cell-free network. Then, for the proposed network in the typical wideband scenario, we formulate the joint precoding design problem to maximize the \ac{wsr} of all users. Finally, inspired by the methods introduced in \cite{Shen'18'1}, we propose a joint precoding framework to solve the formulated problem. Specifically, we decouple the joint precoding design via Lagrangian dual reformulation and fractional programming (FP). Then, by solving the decoupled subproblems alternatively, the \ac{wsr} will converge to a local optimal solution. The simulation results show that \ac{ris}s can improve the cell-free network capacity significantly. 
\par
\begin{figure}[!t]
	\centering
	\includegraphics[width=3.2in]{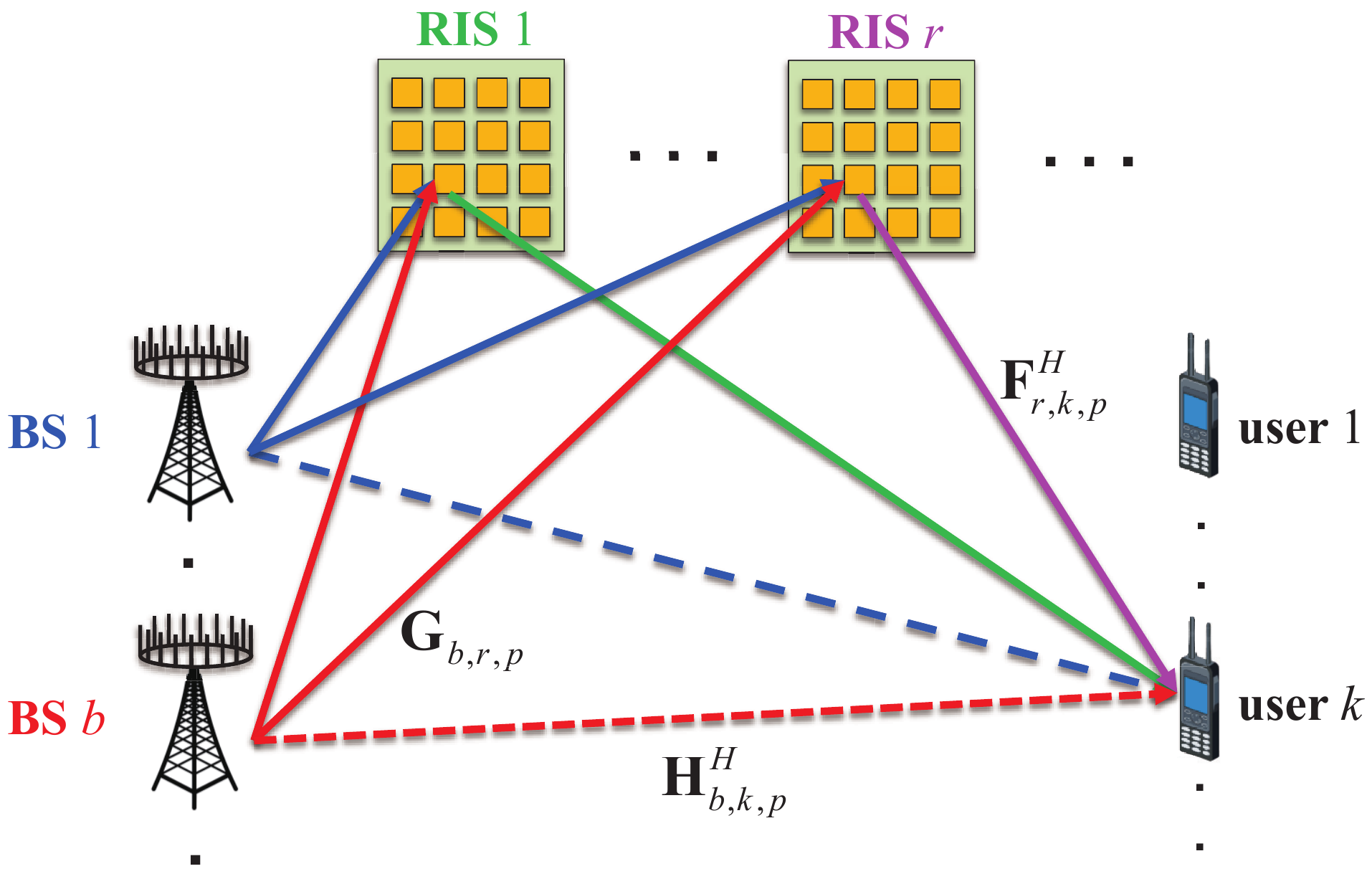}
	\caption{The downlink channels in the wideband RIS-aided cell-free network.}
	\label{img:scenario_show}
\end{figure}
\begin{figure*}[!b]
	\hrulefill
	\setcounter{equation}{5}
	\begin{equation}
	\label{eqn:6}
	\begin{aligned}
	{{\bf y}_{k,p}} =& \sum\limits_{b = 1}^B {{{\bf y}_{b,k,p}}}+{\bf z}_{k,p}  =\sum\limits_{b = 1}^B {\sum\limits_{j = 1}^K {\left({\bf{H}}_{b,k,p}^H + \sum\limits_{r = 1}^R {{\bf{F}}_{r,k,p}^H{\bf{\Theta }}_r^H{{\bf{G}}_{b,r,p}}} \right){{\bf{w}}_{b,p,j}}{s_{p,j}}} }+{\bf z}_{k,p}\\=
	& \underbrace {\sum\limits_{b = 1}^B {\left({\bf{H}}_{b,k,p}^H + \sum\limits_{r = 1}^R {{\bf{F}}_{r,k,p}^H{\bf{\Theta }}_r^H{{\bf{G}}_{b,r,p}}} \right){{\bf{w}}_{b,p,k}}{s_{p,k}}} }_{\text{Desired signal to user $k$}}  + \underbrace {\sum\limits_{b = 1}^B {\sum\limits_{j = 1,j \ne k}^K  \left({\bf{H}}_{b,k,p}^H + \sum\limits_{r = 1}^R {{\bf{F}}_{r,k,p}^H{\bf{\Theta }}_r^H{{\bf{G}}_{b,r,p}}} \right){{\bf{w}}_{b,p,j}}{s_{p,j}}}  }_{\text{ Interference from other users}}+ {{\bf z}_{k,p}}
	\end{aligned}.
	\end{equation}
\end{figure*}
\textit{Notations:} $\mathbb{C}$, $\mathbb{R}$, and $\mathbb{R}^{+}$ denote the set of complex, real, and positive real numbers, respectively; ${[\cdot]^{-1}}$, ${[\cdot]^{*}}$, ${[\cdot]^{T}}$, and ${[\cdot]^{H}}$ denote the inverse, conjugate, transpose, and conjugate-transpose operations, respectively; $\|\cdot\|$ denotes the Euclidean norm of its argument; ${\rm diag}(\cdot)$ denotes the diagonal operation; $\mathfrak{R}\{\cdot\}$ denotes the real part of its argument; $\otimes$ denotes the Kronecker product; $\angle[\cdot]$ denotes the angle of its complex argument; $\mathbf{I}_{L}$ is an $L\times L$ identity matrix, and $\mathbf{0}_{L}$ is an $L\times L$ zero matrix; Finally, $\mathbf{e}_{l}$ is an elementary vector with a one at the $l$-th position, and $\mathbf{1}_{L}$ indicates an $L$-length vector with all elements are 1.
\section{System Model}\label{sec:sys}
To increase the network capacity with low cost and power consumption, in this paper we first propose the concept of RIS-aided cell-free network. Specifically, we consider a wideband scenario as shown in Fig. \ref{img:scenario_show}, where multiple BSs, multiple RISs, multiple users, wideband and MIMO are considered simultaneously. Let ${\cal M} = \left\{ {1, \cdots ,M} \right\}$, ${\cal U} = \left\{ {1, \cdots ,U} \right\}$, ${\cal N} = \left\{ {1, \cdots ,N} \right\}$, ${\cal B} = \left\{ {1, \cdots ,B} \right\}$, ${\cal R} = \left\{ {1, \cdots ,R} \right\}$, ${\cal K} = \left\{ {1, \cdots ,K} \right\}$ and ${\cal P} = \left\{ {1, \cdots ,P} \right\}$ denote the index sets of \ac{bs} antennas, user antennas, \ac{ris} elements, \ac{bs}s, \ac{ris}s, users, and subcarriers, respectively.
\par
In the proposed RIS-aided cell-free network, all \ac{bs}s are synchronized. Let ${{\bf{s}}_{p}} \triangleq {\left[ {{s_{p,1}}, \cdots ,{s_{p,K}}} \right]^T} \in {{\mathbb C}^K}$, where ${s}_{p,k}$ denotes the transmitted symbol to the $k$-th user on the $p$-th subcarrier. We assume that the transmitted symbols have normalized power. In the downlink, the frequency-domain symbol ${s}_{p,k}$ is firstly precoded by the precoding vector ${\bf w}_{b,p,k}\in{{\mathbb C}^{M}}$ at the $b$-th \ac{bs}, so the precoded symbol ${{\bf{x}}_{b,p}}$ at the $b$-th \ac{bs} on the $p$-th subcarrier can be written as
\setcounter{equation}{0}
\begin{equation}
\label{eqn:1}
{{\bf{x}}_{b,p}} = \sum\limits_{k = 1}^K {{{{\bf{w}}_{b,p,k}}{s_{p,k}}}}.
\end{equation}
\par
Thanks to the directional reflection supported by $R$ \ac{ris}s as shown in Fig. \ref{img:scenario_show}, the channel between each \ac{bs} and each user in the proposed RIS-aided cell-free network consists of two parts: the BS-user link and $R$ BS-RIS-user links. Therefore, the equivalent channel ${{\bf{h}}_{b,k,p}^H}$ from the $b$-th \ac{bs} to the $k$-th user on the $p$-th subcarrier can be written as 
\begin{equation}
\label{eqn:2}
{{\bf{h}}_{b,k,p}^H} = \underbrace {{{\bf{H}}_{b,k,p}^H}}_{\text{BS-user link}} + \underbrace {\sum\limits_{r = 1}^R {{{\bf{F}}_{r,k,p}^H{\bf{\Theta }}_r^H{{\bf{G}}_{b,r,p}}}} }_{\text{BS-RIS-user links}},
\end{equation}
where ${{\bf{H}}_{b,k,p}^H}\in\mathbb{C}^{U\times M}$, ${\bf{G}}_{b,r,p}\in\mathbb{C}^{N\times M}$, and ${\bf{F}}_{r,k,p}^H\in\mathbb{C}^{U\times N}$ denote the frequency-domain channel on the subcarrier $p$ from the \ac{bs} $b$ to the user $k$, from the \ac{bs} $b$ to the \ac{ris} $r$, and from the \ac{ris} $r$ to the user $k$, respectively; ${\bf{\Theta }}_r\in\mathbb{C}^{N\times N}$ denotes the phase shift matrix at the \ac{ris} $r$, which is written as
\begin{equation}
\label{eqn:3}
{{\bf{\Theta }}_r} \buildrel \Delta \over = {\mathop{\rm diag}\nolimits} \left( {{\theta _{r,1}}, \cdots ,{\theta _{r,N}}} \right),~\forall r \in \mathcal{R},
\end{equation}
where ${{\theta _{r,n}}}\in{\cal F}$. Note that ${\cal F}$ is the feasible set of the \ac{rc} at \ac{ris}. Here we assume
\begin{equation}
\label{eqn:4}
{{\cal F}} \triangleq \left\{ {\theta _{r,n}}{\Big |} \left| {{\theta _{r,n}}} \right| \le 1\right\} ,~\forall r \in \mathcal{R},~\forall n\in {{\cal N}}.
\end{equation}
\par
After passing through the equivalent channel ${{\bf{h}}_{b,k,p}^H}$ as shown in (\ref{eqn:2}), the signals will be received by the users. Let ${{\bf y}_{b,k,p}}\in\mathbb{C}^U$ denote the baseband frequency-domain signal, which is received by the user $k$ on the subcarrier $p$ from the BS $b$. Since there are $B$ \ac{bs}s serving $K$ users simultaneously, the received signal at user $k$ can be written as (\ref{eqn:6}) at the bottom of this page, where ${\bf z}_{k,p}$ denotes the \ac{awgn} with zero mean and covariance ${\bf{\Xi }}_{k,p}=\sigma^2 {{\bf{I}}_U}$.
\par
Based on the system model above, notice that the received signal ${{\bf y}_{b,k,p}}$ in (\ref{eqn:6}) can be simplified as
\setcounter{equation}{6}
\begin{equation}
\label{eqn:7}
\begin{aligned}
{{\bf{y}}_{k,p}}
\mathop { = }\limits^{(a)} &\sum\limits_{b = 1}^B {\sum\limits_{j = 1}^K {\left( {{\bf{H}}_{b,k,p}^H + {\bf{F}}_{k,p}^H{\bf{\Theta }}^H{{\bf{G}}_{r,p}}} \right){{\bf{w}}_{b,p,j}}{s_{p,j}}} }  + {{\bf{z}}_{k,p}}\\
\mathop { = }\limits^{(b)} &\sum\limits_{j = 1}^K {{\bf{h}}_{k,p}^H{{\bf{w}}_{p,j}}{s_{p,j}}}  + {{\bf{z}}_{k,p}},
\end{aligned}
\end{equation} 
where $(a)$ holds by defining ${\bf{\Theta }} = {\rm diag}({{\bf{\Theta }}_1}, \cdots ,{{\bf{\Theta }}_R})$, ${{\bf{F}}_{k,p}} = {[ {{\bf{F}}_{1,k,p}^T, \cdots ,{\bf{F}}_{R,k,p}^T} ]^T}$, and ${{\bf{G}}_{r,p}} = {\left[ {{\bf{G}}_{1,r,p}^T, \cdots ,{\bf{G}}_{B,r,p}^T} \right]^T}$; $(b)$ holds by defining ${\bf{h}}_{k,p}=[{\bf{h}}_{1,k,p}^T,\cdots,{\bf{h}}_{B,k,p}^T]^T$ and ${\bf{w}}_{p,k}=[{\bf{w}}_{1,p,k}^T,\cdots,{\bf{w}}_{B,p,k}^T]^T$. Then, the \ac{sinr} for the symbol $s_{p,k}$ at the user $k$ on the subcarrier $p$ can be calculated as
\begin{equation}
\label{eqn:8}
\begin{aligned}
&{\gamma _{k,p}} \\
=& {\bf{w}}_{p,k}^H{{\bf{h}}_{k,p}}\!{\left( {\sum\limits_{j = 1,j \ne k}^K \!\!\! {{\bf{h}}_{k,p}^H{{\bf{w}}_{p,j}}{{\left( {{\bf{h}}_{k,p}^H{{\bf{w}}_{p,j}}} \right)}^H}} \!\!+\! {{\bf{\Xi }}_{k,p}}} \right)^{ - 1}}\!\!{\bf{h}}_{k,p}^H{{\bf{w}}_{p,k}}.
\end{aligned}
\end{equation}
Therefore, the \ac{wsr} maximization optimization problem ${\cal P}^{\rm o}$ can be originally formulated as
\begin{subequations}\label{eqn:11}
\begin{align}
\label{eqn:11a}
{\cal P}^{\rm o}:~&\mathop {{\rm{max}}}\limits_{{\bf{\Theta }},{\bf{W}}} ~f({\bf{\Theta }},{\bf{W}})=\sum\limits_{k = 1}^K {\sum\limits_{p = 1}^P {{\eta_k}{{\log }_2}\left( {1 + {\gamma_{k,p}}} \right)} }\\
\label{eqn:11b}
&\,{\rm{s.t.}}~~~C_1:\sum\limits_{k = 1}^K {\sum\limits_{p = 1}^P {\left\| {{{\bf{w}}_{b,p,k}}} \right\|} }  \le {P_{b,\max }},~\forall b \in \mathcal{B},\\
\label{eqn:11c}
&~~~~~~~\,C_2:{{\theta _{r,n}}}\in{\cal F},~\forall r\in{{\cal R}},\forall n\in{{\cal N}},
\end{align}
\end{subequations}
where ${\eta_k}$ denotes the weights of user $k$; ${P_{b,\max }}$ denotes the maximum transmit power of the \ac{bs} $b$, and ${\bf{W}}$ is defined as:
\begin{equation}
\begin{aligned}
{\bf{W}} \!=\! {[{\bf{w}}_{1,1}^T,{\bf{w}}_{1,2}^T, \cdots ,{\bf{w}}_{1,K}^T,{\bf{w}}_{2,1}^T,{\bf{w}}_{2,2}^T, \cdots ,{\bf{w}}_{P,K}^T]^T}.
\end{aligned}
\end{equation}
\section{Proposed Joint Precoding Framework}\label{sec:Alg}
In this section, we present the proposed joint precoding framework to solve the \ac{wsr} optimization problem ${\cal P}^{\rm o}$ in (\ref{eqn:11}).
\subsection{Overview of the proposed joint precoding framework}\label{sec:Alg:p1}
Firstly, to deal with the complexity of sum-logarithms in the \ac{wsr} maximization problem ${\cal P}^{\rm o}$ in (\ref{eqn:11}), based on the multidimensional complex Lagrangian dual reformulation, a method has been proposed in \cite{Shen'18'1} to decouple the logarithms, based on which we have the following \textit{Proposition 1}.
\begin{proposition}\label{proposition:1}
By introducing an auxiliary variable ${\bm \rho }\in \mathbb{R}^{PK}$ with ${\bm \rho } = {[{\rho _{1,1}},\rho _{1,2}^{}, \cdots ,\rho _{1,K},\rho _{2,1}^{},\rho _{2,2}^{}, \cdots ,\rho _{P,K}^{}]^T}$, the original problem ${{\cal P}^{\rm o}}$ in (\ref{eqn:11}) is equivalent to
\begin{equation}
\label{eqn:13}
\begin{aligned}
{\bar {\cal P}}:~~&\mathop {{\rm{max}}}\limits_{{\bf{\Theta }},{\bf{W}},{\bm \rho}}~~~f({\bf{\Theta }},{\bf{W}},{\bm{\rho }})\\
&\,{\rm{s.t.}}~~~C_1:\sum\limits_{k = 1}^K {\sum\limits_{p = 1}^P {\left\| {{{\bf{w}}_{b,p,k}}} \right\|} }  \le {P_{b,\max }},~\forall b \in \mathcal{B},\\
&~~~~~~~\,C_2:{{\theta _{r,n}}}\in{\cal F},~\forall r\in{{\cal R}},\forall n\in{{\cal N}},
\end{aligned}
\end{equation}
where the new objective function $f({\bf{\Theta }},{\bf{W}},{\bm{\rho }})$ is
\begin{equation}
\label{eqn:14}
\begin{aligned}
\!\!f({\bf{\Theta }},{\bf{W}},{\bm{\rho }})\! =\!&\sum\limits_{k = 1}^K {\sum\limits_{p = 1}^P {{\eta _k}{{\log }_2}\left( {1 \!+\! {\rho _{k,p}}} \right)}} \!-\! \sum\limits_{k = 1}^K {\sum\limits_{p = 1}^P {{\eta _k}{\rho _{k,p}}} }  \\&\!+\!  \sum\limits_{k = 1}^K {\sum\limits_{p = 1}^P {{\eta _k}(1 \!+\! {\rho _{k,p}}){f_{k,p}}({\bf{\Theta }},{\bf{W}})} } ,
\end{aligned}
\end{equation}
where the function $f_{k,p}({\bf{\Theta }},{\bf{W}})$ is denoted by
\begin{equation} 
\label{eqn:15}	
\begin{aligned}
&{f_{k,p}}({\bf{\Theta }},{\bf{W}}) =\\& {\bf{w}}_{p,k}^H{{\bf{h}}_{k,p}}\!{\left( {\sum\limits_{j = 1}^K\! {{\bf{h}}_{k,p}^H{{\bf{w}}_{p,j}}{{\left( {{\bf{h}}_{k,p}^H{{\bf{w}}_{p,j}}} \right)}^H}}  \!\!\!+\! {{\bf{\Xi }}_{k,p}}} \right)^{\! \!\!- 1}}\!\!\!\!{\bf{h}}_{k,p}^H{{\bf{w}}_{p,k}}.
\end{aligned}
\end{equation} 	
\end{proposition}
\begin{algorithm}[!b] 
	\caption{Proposed Joint Precoding Framework.} 
	\label{alg:1} 
	\begin{algorithmic}[1] 
		\REQUIRE ~~ 
		The channels ${{\bf{H}}_{b,k,p}},{\bf{G}}_{b,r,p}$ and $ {\bf{F}}_{r,k,p}$ where $\forall b\in {\cal B},k\in {\cal K},p\in {\cal P}$; user weights $\eta_k,~\forall k\in {\cal K}$.
		\ENSURE ~~ 
		Optimized active precoding vector $\bm{W}$; Optimized passive precoding matrix $\bm{\Theta}$; Weighted sum-rate $R_{{\rm sum}}$.	
		\STATE Initialize $\bm{\Theta}$ and $\bm{W}$;
		\WHILE {no convergence of $R_{{\rm sum}}$}
		\STATE Update ${\bm \rho}^{\rm opt}$ by (\ref{eqn:16});
		\STATE Update ${\bm \xi}^{\rm opt}$ by (\ref{eqn:20});
		\STATE Update ${\bf W}^{\rm opt}$ by solving (\ref{eqn:25});
		\STATE Update ${\bm \varpi}^{\rm opt}$ by (\ref{eqn:31});
		\STATE Update ${\bm{\Theta}}^{\rm opt}$ by solving (\ref{eqn:37});
		\STATE Update $R_{{\rm sum}}$ by (\ref{eqn:8}) and (\ref{eqn:11a});
		\ENDWHILE	
		\RETURN ${\bm{\Theta}}$, $\bf{W}$ and $R_{{\rm sum}}$. 
	\end{algorithmic}
\end{algorithm}
\par
Then, after introducing two auxiliary variables ${\bm \xi }$ and $\bm \varpi$, the proposed joint precoding framework to maximize the \ac{wsr} is summarized in {\bf Algorithm 1}, where the optimal solution of $\bf X$ is denoted by ${\bf X}^{\rm opt}$. The method is to fix the other variables and optimize the remain one. The optimal solutions in each step will be introduced in the following three subsections.
\subsection{Fix $({\bf{\Theta}},{\bf{W}})$ and solve ${\bm \rho}^{\rm opt}$}\label{sec:Alg:p2}
Given fixed $({\bf{\Theta }^\star},{\bf{W}^\star})$, the optimal $\bm \rho$ in (\ref{eqn:14}) can be obtained by solving $\partial f / \partial \rho _{k,p}=0$, which can be written as
\begin{equation}
\label{eqn:16}
\begin{aligned}
\rho _{k,p}^{\rm opt} = {\gamma_{k,p}}.
\end{aligned}
\end{equation}
\subsection{Active precoding: fix $({\bf{\Theta }},{\bm{\rho}})$ and solve ${\bf W}^{\rm opt}$} \label{sec:Alg:p3}
In the case of given $({\bf{\Theta}^\star},{\bm{\rho}^\star})$, the equivalent \ac{wsr} maximization problem $\bar{\cal P}$ in (\ref{eqn:13}) can be reformulated as:
\begin{equation}\label{eqn:17}
\begin{aligned}
{{\cal P}_{ \rm active}:}~&\mathop {{\rm{max}}}\limits_{\bf{W}}~g_{1}({\bf{W}})=\sum\limits_{k = 1}^K {\sum\limits_{p = 1}^P {\mu_{k,p}{f_{k,p}}({\bf{\Theta }^\star},{\bf{W}})} } \\
&\,{\rm{s.t.}}~C_1:\sum\limits_{k = 1}^K {\sum\limits_{p = 1}^P {\left\| {{{\bf{w}}_{b,p,k}}} \right\|} }  \le {P_{b,\max }},~\forall b \in \mathcal{B},
\end{aligned}
\end{equation}
where $\mu_{k,p}={\eta _k}(1 + {\rho _{k,p}^\star})$ holds. Inspired by the methods proposed in \cite{Shen'18'1}, we can further obtain \textit{Proposition 2} as below.
\begin{proposition}\label{proposition:2}
	\setcounter{equation}{25}
	\begin{figure*}[!b]
		\hrulefill
		\begin{equation}
		\label{eqn:27}
		\begin{aligned}
		{g_4}({\bf{\Theta }}) = \sum\limits_{k = 1}^K {\sum\limits_{p = 1}^P {\sqrt {{\mu _{k,p}}} {\bf{Q}}_{k,p,k}^H\left( {\bf{\Theta }} \right){{\left( {\sum\limits_{j = 1}^K {{{\bf{Q}}_{k,p,j}}\left( {\bf{\Theta }} \right){\bf{Q}}_{k,p,k}^H\left( {\bf{\Theta }} \right)}  + {{\bf{\Xi }}_{k,p}}} \right)}^{ - 1}}{{\bf{Q}}_{k,p,k}}\left( {\bf{\Theta }} \right)} } .
		\end{aligned}
		\end{equation}
	\end{figure*}
	\setcounter{equation}{15}
With multidimensional complex quadratic transform, by introducing auxiliary variables ${\bm \xi }_{p,k}\in \mathbb{C}^{U}$ and ${\bm \xi }= {[{\bm \xi _{1,1}},\bm\xi _{1,2}, \cdots ,\bm\xi _{1,K},\bm\xi _{2,1},\bm\xi _{2,2}, \cdots ,\bm\xi _{P,K}]}$, the subproblem ${{\cal P}_{ \rm active}}$ in (\ref{eqn:17}) can be further reformulated as
\begin{equation}
\label{eqn:18}
\begin{aligned}
{\bar{\cal P}_{ \rm active}:}~&\mathop {{\rm{max}}}\limits_{\bf{W},{\bm \xi}}~~~g_{2}({\bf{W},{\bm \xi}}) \\
&\,{\rm{s.t.}}~C_1:\sum\limits_{k = 1}^K {\sum\limits_{p = 1}^P {\left\| {{{\bf{w}}_{b,p,k}}} \right\|} }  \le {P_{b,\max }},~\forall b \in \mathcal{B},
\end{aligned}
\end{equation} 
where
\begin{equation}
\label{eqn:19}
\begin{aligned}
&{{g}_2}({\bf{W}},{\bm{\xi }}) = \sum\limits_{k = 1}^K {\sum\limits_{p = 1}^P {2\sqrt {{\mu}_{k,p}}\,\,{\mathfrak{R}} \left\{ {\bm{\xi }}_{k,p}^H{\bf{h}}_{k,p}^H{{\bf{w}}_{p,k}}\right\} } } -
\\&\!\! \sum\limits_{k = 1}^K \!{\sum\limits_{p = 1}^P {{\bm{\xi }}_{k,p}^H\!\left(\sum\limits_{j = 1}^K {{\bf{h}}_{k,p}^H{{\bf{w}}_{p,j}}{{\left( {{\bf{h}}_{k,p}^H{{\bf{w}}_{p,j}}} \right)}^H}}  + {{\bf{\Xi }}_{k,p}} \right){{\bm{\xi }}_{k,p}}} } .
\end{aligned}
\end{equation} 
\end{proposition}
\par
Therefore, we can optimize the variables ${\bf{W}}$ and ${\bm{\xi }}$ in (\ref{eqn:18}) alternatively. The reformulated subproblem ${\bar{\cal P}_{ \rm active}}$ in (\ref{eqn:18}) can be further divided into two subproblems and respectively solved as follows.
\subsubsection{Fix $\bf W$ and solve ${\bm \xi }^{\rm opt}$}
Given fixed $\bf W$, by setting $\partial g_2 / \partial {\bm\xi} _{k,p}$ to zero, the optimal ${\bm \xi }$ can be obtained by
\begin{equation}
\label{eqn:20}
\begin{aligned}
{\bm \xi} _{k,p}^{\rm opt} \!=\!\sqrt {{\mu}_{k,p}} {\left( \sum\limits_{j = 1}^K {{\bf{h}}_{k,p}^H{{\bf{w}}_{p,j}}{{\left( {{\bf{h}}_{k,p}^H{{\bf{w}}_{p,j}}} \right)}^H}}\!\!+ {{\bf{\Xi }}_{k,p}} \right)^{ - 1}}\!\!{\bf{h}}_{k,p}^H{{\bf{w}}_{p,j}}.
\end{aligned}
\end{equation} 
\subsubsection{Fix ${\bm \xi }$ and solve ${\bf W}^{\rm opt}$}
While fixing ${\bm \xi }$ in $\bar{\cal P}_{ \rm active}$ in (\ref{eqn:18}), for simplification and clarity of (\ref{eqn:18}), we can first define
\begin{subequations}\label{eqn:21}
\begin{align}
&{{\bf{a}}_p} = \sum\limits_{k = 1}^K {{\bf{h}}_{k,p}{{\bm{\xi }}_{k,p}}{\bm{\xi }}_{k,p}^H{{\bf{h}}_{k,p}^H}},\\
&{{\bf{A}}_p} = {\bf I}_K \otimes {{\bf{a}}_p},~~~~~~
{\bf v}_{k,p}= {\bm \xi}_{k,p}^H{\bf{h}}_{k,p}^H{{\bf{w}}_{p,k}}.
\end{align}
\end{subequations}
By substituting (\ref{eqn:21}) into $g_2$ in (\ref{eqn:19}), ${{g}_2}$ can be rewritten as
\begin{equation}
\label{eqn:22}
{{g}_3}\left({\bf{W}}\right) =   - {{\bf{W}}^H}{\bf{AW}} +{\mathfrak{R}} \left\{2 {{\bf{V}}^H}{\bf{W}}\right\}- Y,
\end{equation} 
where 
\begin{subequations}
\begin{align}
&{\bf{A}} = {\rm diag}\left( {{{\bf{A}}_1}, \cdots ,{{\bf{A}}_P}} \right),~Y = \sum\limits_{k = 1}^K {\sum\limits_{p = 1}^P {{\bm{\xi }}_{k,p}^H{{\bf{\Xi }}_{k,p}}{{\bm{\xi }}_{k,p}}} },\\
&{\bf{V}} = {[{{\bf v}_{1,1}},{\bf v}_{1,2}^{}, \cdots ,{\bf v}_{1,K}^{},{\bf v}_{2,1}^{},{\bf v}_{2,2}^{}, \cdots ,{\bf v}_{P,K}^{}]^T}.
\end{align}
\end{subequations}
Then, notice that the constraint $C_1$ in (\ref{eqn:18}) can be rewritten as
\begin{equation}
\label{eqn:24}
{\bf{W}}^H{{\bf D}_b}{\bf{W}}  \le {P_{b,\max }},~\forall b \in \mathcal{B},
\end{equation} 
where ${{\bf D}_b}={{\bf{I}}_{PK}} \otimes \left\{ \left({{\bf{e}}_b}{\bf{e}}_b^H\right) \otimes {{\bf{I}}_M}\right\}$ and ${\bf{e}}_b\in\mathbb{R}^{B}$.	
\par
Therefore, by using (\ref{eqn:22}) and (\ref{eqn:24}), the active precoding problem ${\bar{\cal P}_{ \rm active}}$ in (\ref{eqn:18}) can be further simplified as
\begin{equation}
\label{eqn:25}
\begin{aligned}
{\hat{\cal P}_{ \rm active}:}~~&\mathop {{\rm{max}}}\limits_{\bf{W}}~g_{3}({\bf{W}})=- {{\bf{W}}^H}{\bf{AW}} +{\mathfrak{R}} \left\{2 {{\bf{V}}^H}{\bf{W}}\right\} -Y\\
&\,{\rm{s.t.}}~~~~C_1:{\bf{W}}^H{{\bf D}_b}{\bf{W}}  \le {P_{b,\max }},~\forall b \in \mathcal{B}.
\end{aligned}
\end{equation} 
Finally, notice that since the matrices $\bf A$ and ${\bf D}_b$ are all positive semidefinite, the simplified subproblem ${\hat{\cal P}_{ \rm active}}$ in (\ref{eqn:25}) can be directly solved by the standard convex tools such as CVX \cite{cvx}.
\subsection{Passive precoding: fix $({\bm{\rho}},{\bf{W}})$ and solve ${\bf \Theta}^{\rm opt}$}\label{sec:Alg:p4}
Given $({\bm{\rho}}^\star,{\bf{W}}^\star)$, the equivalent problem ${\bar {\cal P}}$ in (\ref{eqn:13}) can be equivalently written as
\begin{equation}
\label{eqn:26}
\begin{aligned}
{{\cal P}_{ \rm passive}:}~~~&\mathop {{\rm{max}}}\limits_{\bf{\Theta}}~~g_{4}({\bf{\Theta}})=\sum\limits_{k = 1}^K {\sum\limits_{p = 1}^P {\mu_{k,p}{f_{k,p}}({\bf{\Theta }},{\bf{W}^\star})} } \\
&~{\rm{s.t.}}~~~C_2:{{\theta _{r,n}}}\in{\cal F},~\forall r\in{{\cal R}},\forall n\in{{\cal N}},
\end{aligned}
\end{equation}
where $\mu_{k,p}={\eta _k}(1 + {\rho _{k,p}^\star})$.
Similarly, to reduce the complexity, we wish to simplify the expression of $g_4$ in (\ref{eqn:26}). Firstly, by defining a new auxiliary function with respect to ${\bf{\Theta}}$ as
\begin{equation}
\label{eqn:28}
{{\bf{Q}}_{k,p,j}}\left( {\bf{\Theta }} \right) = \sum\limits_{b = 1}^B {\left( {{\bf{H}}_{b,k,p}^H + {\bf{F}}_{k,p}^H{{\bf{\Theta }}^H}{{\bf{G}}_{b,p}}} \right){{\bf{w}}_{b,p,j}}},
\end{equation}
we can rewrite $g_4$ in (\ref{eqn:26}) as (\ref{eqn:27}) at the bottom of this page.
However, this subproblem is still hard to solve due to matrix inversion in $f_{k,p}$ in (\ref{eqn:15}). Again, we consider to exploit the multidimensional complex quadratic transform \cite{Shen'18'1} to address this issue by using the following \textit{Proposition 3}.
\begin{proposition}\label{proposition:3}
	With multidimensional complex quadratic transform, by introducing an auxiliary variable ${\bm \varpi }_{p,k}\in \mathbb{C}^{U}$ and ${\bm \varpi }= {[{\bm \varpi _{1,1}},\bm\varpi _{1,2}, \cdots ,\bm\varpi _{1,K},\bm\varpi _{2,1},\bm\varpi _{2,2}, \cdots ,\bm\varpi _{P,K}]}$, the subproblem ${{\cal P}_{ \rm passive}}$ in (\ref{eqn:26}) can be reformulated as
\setcounter{equation}{26}
\begin{subequations}\label{eqn:29}
\begin{align}
\label{eqn:29a}
{\bar{\cal P}_{ \rm passive}:}~~
&\mathop {{\rm{max}}}\limits_{\bf{\Theta}}~~{{g}_5}\left({\bf{\Theta }},{\bm{\varpi }}\right)=\sum\limits_{k = 1}^K {\sum\limits_{p = 1}^P {{g_{k,p}}\left( {\bf{\Theta }},{\bm{\varpi }} \right)} } \\
&~{\rm{s.t.}}~~C_2:{{\theta _{r,n}}}\in{\cal F},~\forall r\in{{\cal R}},\forall n\in{{\cal N}},
\end{align}
\end{subequations}
where	
\begin{equation}
\label{eqn:30}
\begin{aligned}
&{{g}_{k,p}}({\bf{\Theta }},{\bm{\varpi }}) = 2\sqrt {{\mu_{k,p}}}\, {\mathfrak{R}} \left\{ {{\bm{\varpi }}_{k,p}^H{{\bf{Q}}_{k,p,k}}\left( {\bf{\Theta }} \right)} \right\} \\&- {\bm{\varpi }}_{k,p}^H\left( \sum\limits_{j = 1}^K {{{\bf{Q}}_{k,p,j}}\left( {\bf{\Theta }} \right){\bf{Q}}_{k,p,j}^H\left( {\bf{\Theta }} \right)}  + {{\bf{\Xi }}_{k,p}} \right){{\bm{\varpi }}_{k,p}}.
\end{aligned}
\end{equation} 
\end{proposition}
\par
Next, similar to the previous processing, we consider to optimize ${\bf{\Theta }}$ and ${\bm{\varpi }}$ in (\ref{eqn:29}) alternatively as follows.
\subsubsection{Fix $\bm \Theta$ and solve ${\bm \varpi }^{\rm opt}$}
For given fixed $\bm \Theta^\star$, by solving $\partial g_5 / \partial {\bm\varpi} _{k,p}=0$, we can obtain the optimal ${\bm \varpi }$ by
\begin{equation}
\label{eqn:31}
\begin{aligned}
&{\bm{\varpi }}_{k,p}^{\rm opt} =\\& \sqrt {{\mu_{k,p}}} {\left(\sum\limits_{j = 1}^K {{{\bf{Q}}_{k,p,j}}\left( {{{\bf{\Theta }}^{\rm{ \star }}}} \right){\bf{Q}}_{k,p,j}^H\left( {{{\bf{\Theta }}^{\rm{ \star }}}} \right)}  + {{\bf{\Xi }}_{k,p}} \right)^{ \!\!\!- 1}}\!\!{{\bf{Q}}_{k,p,j}}\left( {\bf{\Theta }^\star} \right).
\end{aligned}
\end{equation}
\subsubsection{Fix $\bm \varpi$ and solve ${\bf \Theta}^{\rm opt}$}
While fixing $\bm \varpi$ in $g_5$ in (\ref{eqn:29}), due to the complexity of $\bar{\cal P}_{ \rm passive}$ in (\ref{eqn:29}), we can first define
\begin{subequations}\label{eqn:cg}
\begin{align}
&{\bm{\theta }}={\bm{\Theta }}{\bf 1}_{RN},~~~~
{c_{k,p,j}} = \sum\limits_{b = 1}^B {{\bm\varpi} _{k,p}^H{\bf{H}}_{b,k,p}^H{{\bf{w}}_{b,p,j}}}, \\
&{{\bf{g}}_{k,p,j}} = \sum\limits_{b = 1}^B {{\rm diag}\left( {{\bm\varpi} _{k,p}^H{\bf{F}}_{k,p}^H} \right){{\bf{G}}_{b,p}}{{\bf{w}}_{b,p,j}}}.
\end{align}
\end{subequations}
Then, by substituting (\ref{eqn:28}) and (\ref{eqn:cg}) into $g_5$ in (\ref{eqn:29a}), $g_5$ can be rewritten as
\begin{equation}
\label{eqn:35}
\begin{aligned}
{{g}_6}({\bm{\Theta}}) =   - {{\bm{\theta}}^H}{\bf{	\Lambda}}{\bm \theta} +{\mathfrak{R}} \left\{2 {{\bm{\theta}}^H}{\bm{\nu}}\right\}- 	\zeta,
\end{aligned}
\end{equation}
where
\begin{subequations}
	\begin{align}
	{\bm\Lambda } =& \sum\limits_{k = 1}^K {\sum\limits_{p = 1}^P {\sum\limits_{j = 1}^K {{{\bf{g}}_{k,p,j}}{\bf{g}}_{k,p,j}^H} } }, \\
	{\bm\nu } =& \sum\limits_{k = 1}^K {\sum\limits_{p = 1}^P {\sqrt {{\mu _{k,p}}}{{\bf{g}}_{k,p,k}}} }  \!-\! \sum\limits_{k = 1}^K {\sum\limits_{p = 1}^P {\sum\limits_{j = 1}^K {c_{k,p,j}^*{{\bf{g}}_{k,p,j}}} } }, \\
	{\zeta}  =&  \sum\limits_{k = 1}^K {\sum\limits_{p = 1}^P {\sum\limits_{j = 1}^K {{{\left| {{c_{k,p,j}}} \right|}^2}} } } 
	+ \sum\limits_{k = 1}^K {\sum\limits_{p = 1}^P {{\bm\varpi} _{k,p}^H{{\bf{\Xi }}_{k,p}}{{\bm\varpi} _{k,p}}} } 
	\\&-2\sum\limits_{k = 1}^K {\sum\limits_{p = 1}^P {\sqrt {{\mu _{k,p}}}}~{\mathfrak{R}}\left\{{c_{k,p,k}} \right\}}.
	\end{align}
\end{subequations}
Therefore, the reformulated passive precoding subproblem ${\bar{\cal P}_{ \rm passive}}$ in (\ref{eqn:29}) can be further simplified as
\begin{equation}
\label{eqn:37}
\begin{aligned}
{\hat{\cal P}_{ \rm passive}:}~~&\mathop {{\rm{max}}}\limits_{\bf{\Theta}}~~{{g}_6}({\bm{\Theta}}) =   - {{\bm{\theta}}^H}{\bf{	\Lambda}}{\bm \theta} +{\mathfrak{R}} \left\{2 {{\bm{\theta}}^H}{\bm{\nu}}\right\} -\zeta\\
&\,{\rm{s.t.}}~~~C_2:{{\theta _{r,j}}}\in{\cal F},~\forall r\in{{\cal R}},\forall j\in{{\cal N}},
\end{aligned}
\end{equation}  
which is similar to those in \cite{Guo'19,Wu'19}. Since the matrix ${\bm\Lambda }$ and the constraint $C_2$ are positive semidefinite, ${\bf \Theta}^{\rm opt}$ can be directly obtained by the standard convex tools such as CVX \cite{cvx}. 
\vspace{2mm}
\section{Simulation Results}\label{sec:NSR}
\begin{figure}[!t]
	\centering
	\includegraphics[width=3in]{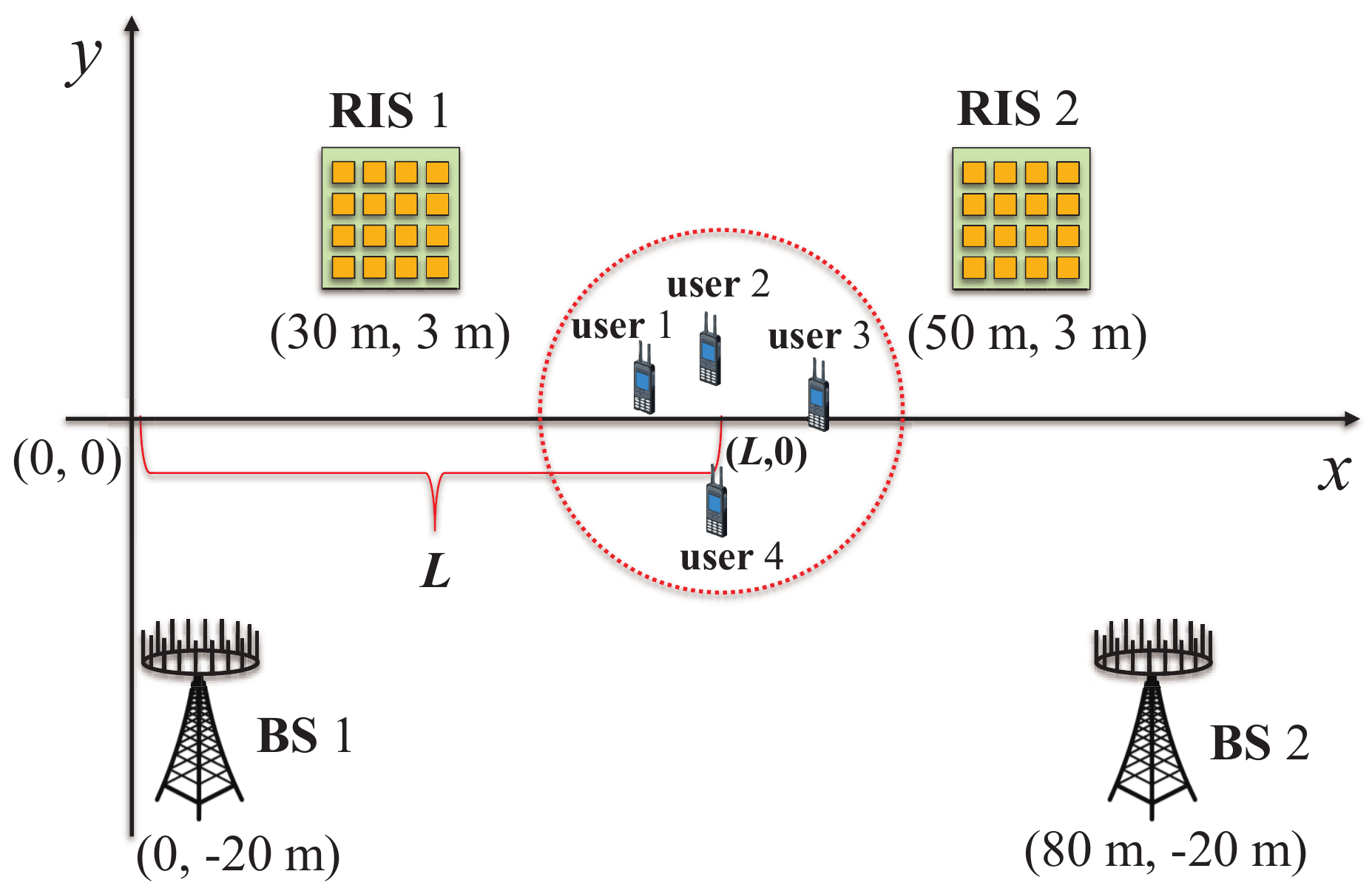}
	\caption{Scenario where two \ac{bs}s assisted by two \ac{ris}s serve four users.}
	\label{img:simulation_scenario_1}
\end{figure}
For the simulation of the proposed RIS-aided cell-free network, we consider a RIS-aided cell-free system with the topology as shown in Fig. \ref{img:simulation_scenario_1}, where four users are randomly distributed in a circle centered at $(L,0)$ with radius $\rm 1\,m$. The other system parameters are set as follows: $P_{b,\max}=0$ dB, $M=8$, $U=2$, $N=32$, $P=6$, $\eta_{k}=1$, and ${\sigma ^2} =  - 120$ dBm. The signal attenuation is set as $30$ dB at a reference distance $1$ m for the BS-user link and $40$ dB for the BS-RIS-user link, respectively. The path loss exponent of the BS-RIS link, the RIS-user link, and the BS-user link is set as $2$, $2$, and $3$, respectively. We also assume that, the RIS-user link and the BS-user link satisfy Rayleigh fading, while the BS-RIS link satisfies LoS fading. $\bm \Theta$ is initialized by random values in ${\cal F}$, and $\bf W$ is initialized by identical power and random phases.
\par
\begin{figure}[!t]
	\centering
	\includegraphics[width=3in]{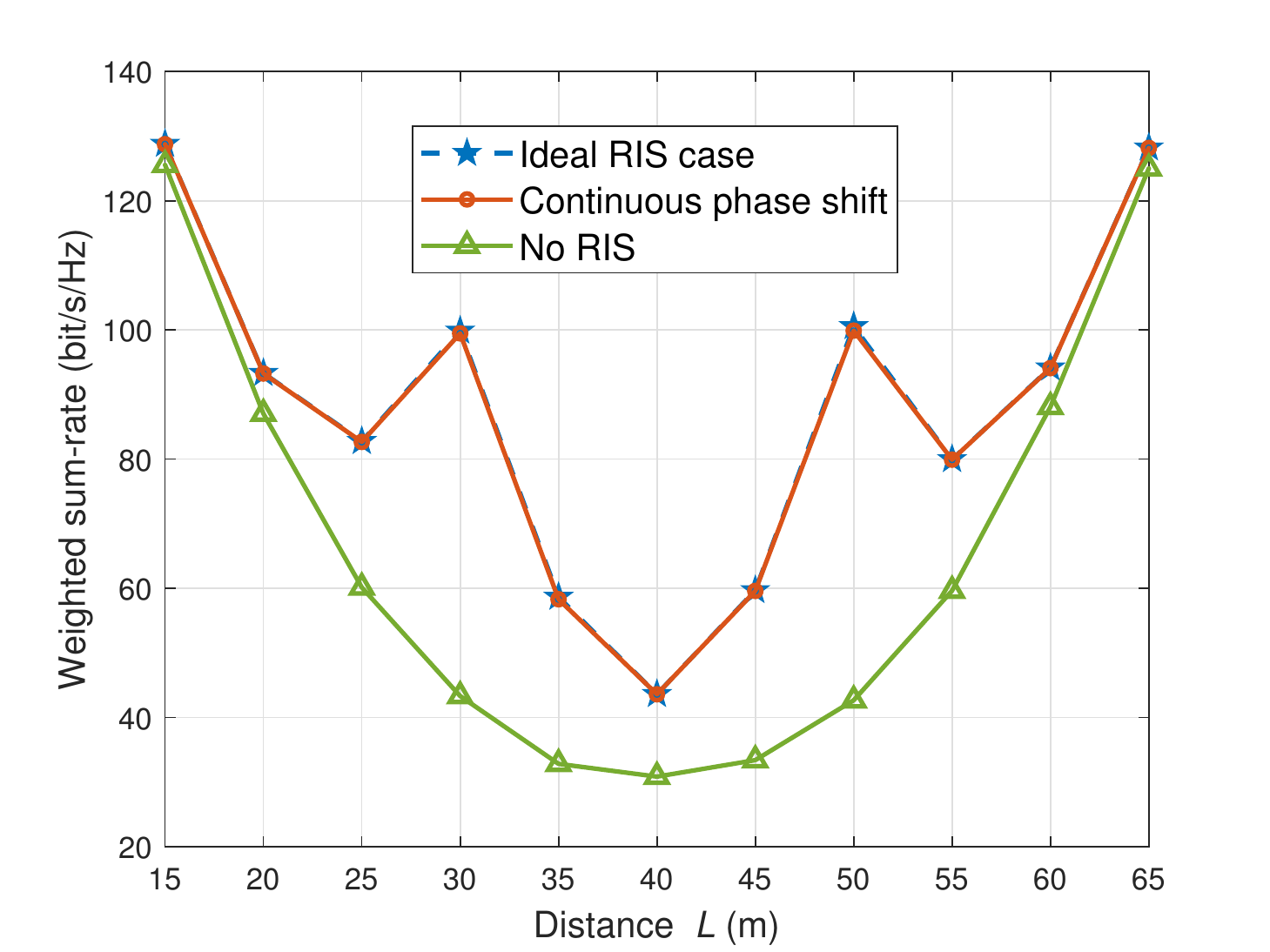}
	\caption{Weighted sum-rate against the distance $L$.}
	\label{img:simulation_1}
\end{figure}

Fig. \ref{img:simulation_1} shows the \ac{wsr} against the distance $L$, where the conventional cell-free network without \ac{ris} (which is denote by “No RIS”) is considered as benchmark. Specifically, the “Ideal RIS case” denotes the case where the shift of RIS elements satisfy (\ref{eqn:4}), while “Continuous phase shift” denotes the case where ${{\theta _{r,n}}}\in\{ {\theta _{r,n}}{ |} \left| {{\theta _{r,n}}} \right| = 1\}$, which is obtained by applying approximation operation on ${\cal F}$. To the best of our knowledge, the wideband \ac{wsr} maximization problem in conventional cell-free network has not been investigated in the literature. Therefore, the curve “No RIS” is achieved by the proposed framework with the setting $R=0$. 

From Fig. \ref{img:simulation_1}, we can see that, as the users move away from the \ac{bs}s, the \ac{wsr} of the users decreases rapidly. However, for the proposed RIS-aided cell-free network, we can see two obvious peaks at $L=30\,{\rm m}$ and $L=50\,{\rm m}$, which indicates that the \ac{wsr} rises when the users approach one of the two \ac{ris}s, since the users can receive strong signals reflected from the \ac{ris}s. Thus, we can conclude that the network capacity can be substantially increased by deploying RISs in the network, and the signal coverage can be accordingly extended. Limited by the length of the article, please refer to \cite{Zijian} for more simulation results and insights.
\section{Conclusions}\label{sec:con}
In this paper, we first propose the concept of RIS-aided cell-free network, which aims to improve the network capacity with low cost and power consumption. Then, for the proposed network, in the typical wideband scenario, we formulate the joint precoding design problem to maximize the weighted sum-rate of all users. Finally, we propose a joint precoding framework to solve this problem. Since most of the considered scenarios in existing works are special cases of the general scenario in this paper, the proposed joint precoding framework can also serve as a general solution to maximize the capacity in most of existing RIS-aided scenarios. The simulation results demonstrate that, the proposed RIS-aided cell-free network can realize higher capacity than the conventional cell-free network.
\section*{Acknowledgment}
This work was supported by the National Science and Technology Major Project of China under Grant 2018ZX03001004-003 and the National Natural Science Foundation of China for Outstanding Young Scholars under Grant 61722109.
\footnotesize
\balance 
\bibliographystyle{IEEEtran}
\bibliography{IEEEabrv,reference}
\end{document}